\documentclass[12pt]{article}
\usepackage{amssymb, amsmath}
\usepackage{amsfonts}
\usepackage{color}
\usepackage{mathrsfs}
\usepackage{cite}

\newcommand{\ret}{\nonumber \\}

\newcommand{\Section}[1]%
{\section{#1}\setcounter{equation}{0}%
\setcounter{theorem}{0}}
\newtheorem{theorem}{Theorem}

{\par\noindent{\em #1:\ }}%
{~\rule{2mm}{2mm}\par\bigskip}
\def\re{\mathbb{R}}

\def\ze{\mathbb{Z}}

\begin{document}
\newpage\thispagestyle{empty}
{\topskip 2cm
\begin{center}
{\Large\bf $\pi$ Flux Phase and Superconductivity\\  
\bigskip
for Lattice Fermions Coupled to Classical Gauge Fields\\}
\bigskip\bigskip
{\Large Tohru Koma\\}
\bigskip\medskip
{\small Department of Physics, Gakushuin University (retired), 
Mejiro, Toshima-ku, Tokyo 171-8588, JAPAN\\}
\smallskip

\end{center}
\vfil
\noindent
{\bf Abstract:} \\
We study superconducting lattice fermions coupled to classical gauge fields. Namely, without the gauge fields, 
the lattice fermions show superconducting long-range order. In the previous paper, the existence of the long-range order 
was proved by the present author. This paper is the continuing part of the previous one.  
More precisely, the interactions between fermions were assumed to be a Bardeen-Cooper-Schrieffer-type pairing  
which is a nearest-neighbour two-body interaction on the hypercubic lattice $\ze^d$ with the dimension $d\ge 3$. 
In the present paper, we deal with the corresponding gauged model by introducing classical U(1) gauge fields. 
We prove that a certain configuration of gauge fields which yields the $\pi$ flux for all the plaquettes in the lattice 
minimizes the ground-state energy of the fermion system. Since this configuration of the gauge fields exactly coincides 
with the hopping amplitudes of the Hamiltonian of the previous paper, the ground state of the whole system shows 
superconducting long-range order. Any configuration of gauge fields which is gauge-equivalent to this special configuration of 
the gauge fields also yields another ground state. However, if the Cooper pair correlations are averaged over 
the gauge-equivalent class, then they are vanishing except for the on-site correlation because the Cooper pair correlations 
are not necessarily gauge invariant. On the other hand, the string Cooper pair correlations which are gauge invariant 
always show the same long-range order for any configuration of gauge fields in the gauge-equivalent class.    
\par\noindent
\bigskip
\hrule
\bigskip


\vfil}\newpage
\tableofcontents
\newpage
\Section{Introduction}

In the previous paper \cite{Koma4}, the present author proved the existence of the superconducting long-range order 
in a lattice fermion system with nearest-neighbour two-body interactions 
on the hypercubic lattice $\ze^d$ with the dimension $d\ge 3$. 
These interactions are a Bardeen-Cooper-Schrieffer-type pairing \cite{BCS}. 
In addition, as proved in the paper, the long-range order implies the existence of the U(1) symmetry 
breaking ground state \cite{KomaTasaki1} and the emergence of the Nambu-Goldstone mode \cite{Nambu,NJL,Goldstone,GSW} 
above the ground state \cite{Koma1}. 
However, there is no experimental evidence of the U(1) symmetry breaking in superconductors. 
On the other hand, when gauge fields are introduced into the system, the concept of gauge invariance plays an important role 
for treating physical observables. Actually, since the Cooper pair itself is not gauge invariant, 
it might not be a physical observable.   
 
In the present paper, we deal with this issue by introducing classical U(1) gauge fields into the superconducting fermion model 
which we considered in the previous paper \cite{Koma4}. 
Namely, we study the superconducting lattice fermions coupled to the classical gauge fields. 
Thus, the present paper is the continuing part of the previous one.  
We prove that a certain configuration of gauge fields which yields the $\pi$ flux \cite{LiebFlux,MN} 
for all the plaquettes in the lattice 
minimizes the ground-state energy of the fermion system. The precise statement will be given by Theorem~\ref{maintheorem} 
in Section~\ref{Sec:RPreal}. Since this configuration of the gauge fields exactly coincides 
with the hopping amplitudes of the Hamiltonian of the previous paper, the ground state of the whole system shows 
superconducting long-range order. Any configuration of gauge fields which is gauge-equivalent to this special configuration of 
the gauge fields also yields another ground state. However, if the Cooper pair correlations are averaged over 
the gauge-equivalent class, then they are vanishing except for the on-site correlation because the Cooper pair correlations 
are not necessarily gauge invariant. On the other hand, the string Cooper pair correlations which are gauge invariant 
always show the same long-range order for any configuration of gauge fields in the gauge-equivalent class.    
These arguments will be given in Section~\ref{Sec:LRO}.

\Section{Model} 

We first describe our model. Consider a $d$-dimensional finite hypercubic lattice, 
\begin{equation}
\label{Lambda}
\Lambda:=\{-L+1,-L+2,\ldots,-1,0,1,\ldots,L-1,L\}^d\subset\mathbb{Z}^d,
\end{equation}
with a positive integer $L$ and $d\ge 3$. 
We write $x=(x^{(1)},x^{(2)},\ldots,x^{(d)})\in\ze^d$ for the $d$-dimensional coordinates, and 
we impose the periodic boundary condition, $x^{(i)}=L+1\equiv -L+1$, $i=1,2,\ldots,d$, for the lattice $\Lambda$. 
For each lattice site $x\in\Lambda$, 
we introduce a fermion operator $a_{x,\sigma}$ with spin degrees of freedom $\sigma=\uparrow,\downarrow$. 
The fermion operators satisfy the anticommutation relations, 
\begin{equation}
\{a_{x,\sigma},a_{x',\sigma'}^\dagger\}=\delta_{x,x'}\delta_{\sigma,\sigma'}\quad 
\mbox{and}\quad \{a_{x,\sigma},a_{x',\sigma'}\}=0,
\end{equation}
for $x,x'\in\Lambda$ and $\sigma,\sigma'=\uparrow,\downarrow$. The number operator is given by 
\begin{equation}
n_{x,\sigma}:=a_{x,\sigma}^\dagger a_{x,\sigma}
\end{equation}
for $x\in\Lambda$ and $\sigma=\uparrow,\downarrow$. 
The hopping Hamiltonian on the lattice $\Lambda$ for fermions is given by 
\begin{equation}
\label{Hhop}
H_{\rm hop}^{(\Lambda)}(A):=\kappa \sum_{\sigma=\uparrow,\downarrow}\sum_{\substack{\{x,y\}\subset\Lambda\\ : |x-y|=1}}
(a_{x,\sigma}^\dagger e^{iA_{x,y}}a_{y,\sigma}+ a_{y,\sigma}^\dagger e^{-iA_{x,y}}a_{x,\sigma}). 
\end{equation}
Here, $\kappa\in\re$ is the hopping amplitude, and $A_{x,y}\in(-\pi,\pi]$ are classical gauge fields.  
We assume that the interactions between fermions are described by Bardeen-Cooper-Schrieffer-type pairing \cite{BCS},
\begin{equation}
\label{Hint}
H_{\rm int}^{(\Lambda)}(A):=-g \sum_{\substack{\{x,y\}\subset\Lambda\\ : |x-y|=1}}
(e^{2iA_{x,y}}a_{x,\uparrow}^\dagger a_{x,\downarrow}^\dagger a_{y,\downarrow}a_{y,\uparrow}+
e^{-2iA_{x,y}}a_{y,\uparrow}^\dagger a_{y,\downarrow}^\dagger a_{x,\downarrow}a_{x,\uparrow}), 
\end{equation}
with the coupling constant $g\ge 0$. Then, the full Hamiltonian that we consider is given by  
\begin{equation}
H^{(\Lambda)}(A):=H_{\rm hop}^{(\Lambda)}(A)+H_{\rm int}^{(\Lambda)}(A)
+K\sum_{\wp\subset\Lambda}\cos F_\wp,
\end{equation}
where $\wp$ denotes a plaquette (unit square cell), which consists of four bounds, i.e., 
$\wp=\{\{x,y\},\{y,z\},\{z,u\},\{u,x\}\}$ with $x,y,z,u\in\Lambda$ satisfying $|x-y|=1$, and $F_\wp$ is given by 
\begin{equation}
\label{F}
F_\wp=A_{x,y}+A_{y,z}+A_{z,u}+A_{u,x}.  
\end{equation}
In the following, we assume $K>0$. 

\Section{$\pi$ flux phase}

Although the above assumption $K>0$ is not the usual convention, our idea is as follows: 
We change the variables $A_{x,y}$ to $\tilde{A}_{x,y}$ by 
\begin{equation}
\label{def:tildeA}
e^{iA_{x,x+e_i}}=i(-1)^{\theta_i(x)}e^{i\tilde{A}_{x,x+e_i}}
\end{equation}
for $i=1,2,\ldots,d$, where $\theta_i(x)$ is given by 
\begin{equation}
\label{theta1}
\theta_1(x):=
\begin{cases}
0 & \mbox{for \ } x^{(1)}\ne L;\\
1 & \mbox{for \ } x^{(1)}=L,
\end{cases}
\end{equation}
and 
\begin{equation}
\label{thetai}
\theta_i(x):= 
\begin{cases}
{x^{(1)}+\cdots+x^{(i-1)}} & \mbox{for \ } x^{(i)}\ne L;\\
{x^{(1)}+\cdots+x^{(i-1)}}+1 & \mbox{for \ } x^{(i)}=L,
\end{cases}
\end{equation}
for $i=2,3,\ldots,d$, and $e_i$ is the unit vector whose $i$-th component is $1$. 
Here, we stress that the condition about $x^{(i)}=L$ yields an anti-periodic boundary condition for the hopping Hamiltonian. 
[See (\ref{tildeHhop}) below.] The condition is useful for applying 
the method of reflection positivity \cite{FSS,DLS,FILS1,FILS,KLS,KLS2,JP} later. 
When the gauge field $\tilde{A}$ is equal to zero, i.e., $\tilde{A}_{x,y}=0$ for all the bonds $\{x,y\}$, 
the present Hamiltonian for fermions is the same as that in the previous paper \cite{Koma4}. See (\ref{tildeHint}), (\ref{tildeHhop}) 
and (\ref{fullbarHtildeA}) below. Besides, all the plaquettes have the same $\pi$ flux \cite{LiebFlux,MN} for $\tilde{A}=0$. 
In the rest of this section, we will show these statements. 

Consider a plaquette $\wp=\{\{x,x+e_i\},\{x+e_i,x+e_i+e_j\},\{x+e_i+e_j,x+e_j\},\{x+e_j,x\}\}$ for $i\ne j$. 
The relation (\ref{def:tildeA}) is equivalent to 
\begin{equation}
A_{x,x+e_i}=\frac{\pi}{2}+\pi\theta_i(x)+\tilde{A}_{x,x+e_i}.
\end{equation}
Similarly, 
\begin{equation}
A_{x+e_i+e_j,x+e_j}=-\frac{\pi}{2}-\pi\theta_i(x+e_j)+\tilde{A}_{x+e_i+e_j,x+e_j},
\end{equation}
where we have used $A_{y,x}=-A_{x,y}$. By adding both sides of these equations, one has  
\begin{equation}
A_{x,x+e_i}+A_{x+e_i+e_j,x+e_j}=\tilde{A}_{x,x+e_i}+\tilde{A}_{x+e_i+e_j,x+e_j}+\pi[\theta_i(x)-\theta_i(x+e_j)].
\end{equation}
The third term in the right-hand side is written 
\begin{equation}
\pi[\theta_i(x)-\theta_i(x+e_j)]=
\begin{cases}
0 \ \ \mbox{mod \;} 2\pi &  \mbox{for \ } j > i; \\
\pi \ \ \mbox{mod \;} 2\pi & \mbox{for \ } j < i,
\end{cases}
\end{equation}
{from} the definitions (\ref{theta1}) and (\ref{thetai}) of $\theta_i(x)$. 
In the same way, one has 
\begin{equation}
A_{x+e_i,x+e_i+e_j}+A_{x+e_j,x}=\tilde{A}_{x+e_i,x+e_i+e_j}+\tilde{A}_{x+e_j,x}
+\pi[\theta_j(x+e_i)-\theta_j(x)]
\end{equation}
and 
\begin{equation}
\pi[\theta_j(x+e_i)-\theta_j(x)]=
\begin{cases}
0 \ \ \mbox{mod \;} 2\pi & \mbox{for \ }i>j;\\
\pi \ \ \mbox{mod \;} 2\pi & \mbox{for \ } i<j. 
\end{cases}
\end{equation}
Substituting these into the field strength $F_\wp$ of (\ref{F}), we have 
\begin{equation}
F_\wp=\tilde{F}_\wp +\pi,
\end{equation}
where $\tilde{F}_\wp$ is the field strength for the gauge fields $\tilde{A}_{x,y}$. 
Therefore, one has 
\begin{equation}
\cos F_\wp=-\cos \tilde{F}_\wp.
\end{equation}
The minus sign in the right-hand side yields the usual convention for the Hamiltonian, and the hopping amplitudes of 
the hopping Hamiltonian $H_{\rm hop}^{(\Lambda)}$ change to those in the $\pi$-flux phase \cite{LiebFlux,MN} 
when $\tilde{A}_{x,y}=0$ for all $x,y$, i.e., all the plaquettes have the same $\pi$ flux for $\tilde{A}=0$. 
In the following, we will treat the gauge fields $\tilde{A}_{x,y}$ as perturbation from the $\pi$ flux 
because we can expect that the $\pi$ flux minimizes the energy of the system.  

{From} (\ref{def:tildeA}), one has 
$$
e^{2iA_{x,y}}=-e^{2i\tilde{A}_{x,y}}.
$$
Substituting this into the expression (\ref{Hint}) of the interaction Hamiltonian $H_{\rm int}^{(\Lambda)}(A)$, one has  
\begin{equation}
\label{tildeHint}
H_{\rm int}^{(\Lambda)}(A)=g \sum_{\substack{\{x,y\}\subset\Lambda\\ : |x-y|=1}}
(e^{2i\tilde{A}_{x,y}}a_{x,\uparrow}^\dagger a_{x,\downarrow}^\dagger a_{y,\downarrow}a_{y,\uparrow}+
e^{-2i\tilde{A}_{x,y}}a_{y,\uparrow}^\dagger a_{y,\downarrow}^\dagger a_{x,\downarrow}a_{x,\uparrow})
=:\bar{H}_{\rm int}^{(\Lambda)}(\tilde{A}).  
\end{equation}
Similarly, the hopping Hamiltonian $H_{\rm hop}^{(\Lambda)}(A)$ of (\ref{Hhop}) is written 
\begin{equation}
\label{tildeHhop}
H_{\rm hop}^{(\Lambda)}(A)=i\kappa\sum_{\sigma=\uparrow,\downarrow}\sum_{x\in\Lambda}\sum_{i=1}^d 
(-1)^{\theta_i(x)}(a_{x,\sigma}^\dagger e^{i\tilde{A}_{x,x+e_i}}a_{x+e_i,\sigma}
-a_{x+e_i,\sigma}^\dagger e^{-i\tilde{A}_{x,x+e_i}}a_{x,\sigma})=:\bar{H}_{\rm hop}^{(\Lambda)}(\tilde{A}). 
\end{equation}
When $\tilde{A}=0$, these two Hamiltonians are the same as those in \cite{Koma4}. 
In the following, we will deal with the full Hamiltonian which is given by  
\begin{equation}
\label{fullbarHtildeA}
\bar{H}^{(\Lambda)}(\tilde{A}):=\bar{H}_{\rm hop}^{(\Lambda)}(\tilde{A})+\bar{H}_{\rm int}^{(\Lambda)}(\tilde{A})
-K\sum_{\wp\subset\Lambda}\cos \tilde{F}_\wp.
\end{equation}
Our goal is to prove that the gauge fields $\tilde{A}$ given by the gauge equivalent class $\tilde{A}=0$ minimize 
the ground-state energy.   

\Section{Reflection positivity --- spin space}

In this section, we will prove the inequality (\ref{Zbound}) below by relying on the reflection positivity 
about the spin space \cite{Lieb,KuboKishi}. In this upper bound, the gauge field $\tilde{A}$ of the interaction Hamiltonian 
can be taken to be equal to zero.  

We define an anti-linear map $\vartheta_{\rm spin}$ by \cite{JP,Koma4}
\begin{equation}
\vartheta_{\rm spin}(a_{x,\uparrow})=a_{x,\downarrow}\quad \mbox{and}\quad 
\vartheta_{\rm spin}(a_{x,\uparrow}^\dagger)=a_{x,\downarrow}^\dagger
\end{equation}
for $x\in\Lambda$. Here, we have used the fact that the fermion operators $a_{x,\sigma}$ have a real representation. 
For operators $\mathcal{A},\mathcal{B}$, 
\begin{equation}
\vartheta(\mathcal{A}\mathcal{B})=\vartheta(\mathcal{A})\vartheta(\mathcal{B})
\quad \mbox{and}\quad \vartheta(\mathcal{A})^\dagger=\vartheta(\mathcal{A}^\dagger). 
\end{equation}
We also introduce a unitary transformation, 
\begin{equation}
U_{{\rm odd},\pi/2}^{(\Lambda)}:=\prod_{\sigma=\uparrow,\downarrow}\prod_{x\in\Lambda_{\rm odd}}
e^{(i\pi/2)n_{x,\sigma}}, 
\end{equation}
where 
\begin{equation}
\Lambda_{\rm odd}:=\{x\in\Lambda \; | \; x^{(1)}+\cdots+x^{(d)}={\rm odd}\}.
\end{equation}
We write 
\begin{equation}
\label{checkHhop}
\check{H}_{\rm hop}^{(\Lambda)}(\tilde{A})
:=(U_{{\rm odd},\pi/2}^{(\Lambda)})^\dagger\bar{H}_{\rm hop}^{(\Lambda)}(\tilde{A})U_{{\rm odd},\pi/2}^{(\Lambda)}.
\end{equation}
Note that 
\begin{eqnarray}
& &(U_{{\rm odd},\pi/2}^{(\Lambda)})^\dagger i(a_{x,\sigma}^\dagger e^{i\tilde{A}_{x,x+e_i}}a_{x+e_i,\sigma}
-a_{x+e_i,\sigma}^\dagger e^{-i\tilde{A}_{x,x+e_i}}a_{x,\sigma})U_{{\rm odd},\pi/2}^{(\Lambda)}\ret
&=&
\begin{cases}
(a_{x,\sigma}^\dagger e^{i\tilde{A}_{x,x+e_i}}a_{x+e_i,\sigma}
+a_{x+e_i,\sigma}^\dagger e^{-i\tilde{A}_{x,x+e_i}}a_{x,\sigma}) & \mbox{for \ } x\in\Lambda_{\rm odd};\\
-(a_{x,\sigma}^\dagger e^{i\tilde{A}_{x,x+e_i}}a_{x+e_i,\sigma}
+a_{x+e_i,\sigma}^\dagger e^{-i\tilde{A}_{x,x+e_i}}a_{x,\sigma}) & \mbox{for \ } x+e_i\in\Lambda_{\rm odd}.
\end{cases}
\end{eqnarray}
{From} this expression, the Hamiltonian $\check{H}_{\rm hop}^{(\Lambda)}(\tilde{A})$ can be written 
\begin{equation}
\label{vheckHhop}
\check{H}_{\rm hop}^{(\Lambda)}(\tilde{A})=\check{H}_{{\rm hop},\uparrow}^{(\Lambda)}(\tilde{A})
+\check{H}_{{\rm hop},\downarrow}^{(\Lambda)}(\tilde{A})
=\check{H}_{{\rm hop},\uparrow}^{(\Lambda)}(\tilde{A})
+\vartheta_{\rm spin}(\check{H}_{{\rm hop},\uparrow}^{(\Lambda)}(-\tilde{A})). 
\end{equation}
We also have 
\begin{eqnarray}
\label{checkHint}
\check{H}_{\rm int}^{(\Lambda)}(\tilde{A})
&:=&(U_{{\rm odd},\pi/2}^{(\Lambda)})^\dagger\bar{H}_{\rm int}^{(\Lambda)}(\tilde{A})U_{{\rm odd},\pi/2}^{(\Lambda)}\ret
&=&-g \sum_{|x-y|=1}(e^{2i\tilde{A}_{x,y}}a_{x,\uparrow}^\dagger a_{x,\downarrow}^\dagger a_{y,\downarrow}a_{y,\uparrow}
+e^{-2i\tilde{A}_{x,y}}a_{y,\uparrow}^\dagger a_{y,\downarrow}^\dagger a_{x,\downarrow}a_{x,\uparrow})\ret
&=&-g \sum_{|x-y|=1}[e^{2i\tilde{A}_{x,y}}a_{x,\uparrow}^\dagger a_{y,\uparrow}
\vartheta_{\rm spin}(a_{x,\uparrow}^\dagger a_{y,\uparrow})+e^{-2i\tilde{A}_{x,y}}a_{y,\uparrow}^\dagger a_{x,\uparrow}
\vartheta_{\rm spin}(a_{y,\uparrow}^\dagger a_{x,\uparrow})]. \ret
\end{eqnarray}

We define 
\begin{equation}
I_{M,{\rm spin}}:=\Bigl[1-\frac{\beta }{M}\check{H}_{\rm int}^{(\Lambda)}(\tilde{A})\Bigr]
\exp\bigl[-\frac{\beta}{M}\check{H}_{{\rm hop},\uparrow}^{(\Lambda)}(\tilde{A})\bigr]
\exp\bigl[-\frac{\beta}{M}\vartheta_{\rm spin}(\check{H}_{{\rm hop},\uparrow}^{(\Lambda)}(-\tilde{A}))\bigr]
\end{equation}
for an integer $M>0$, where $\beta$ is the inverse temperature. Then, from (\ref{vheckHhop}) and (\ref{checkHint}), 
we have 
\begin{equation}
e^{-\beta\check{H}^{(\Lambda)}(\tilde{A})}=\lim_{M\nearrow\infty}(I_{M,{\rm spin}})^M,
\end{equation}
where we have written 
\begin{equation}
\check{H}^{(\Lambda)}(\tilde{A}):=\check{H}_{\rm hop}^{(\Lambda)}(\tilde{A})
+\check{H}_{\rm int}^{(\Lambda)}(\tilde{A}).
\end{equation}
Note that 
\begin{eqnarray}
\label{IMspinM}
(I_{M,{\rm spin}})^M&=&\Bigl[1-\frac{\beta }{M}\check{H}_{\rm int}^{(\Lambda)}(\tilde{A})\Bigr]
\exp\bigl[-\frac{\beta}{M}\check{H}_{{\rm hop},\uparrow}^{(\Lambda)}(\tilde{A})\bigr]
\exp\bigl[-\frac{\beta}{M}\vartheta_{\rm spin}(\check{H}_{{\rm hop},\uparrow}^{(\Lambda)}(-\tilde{A}))\bigr]\ret
&\times&\cdots \times\Bigl[1-\frac{\beta }{M}\check{H}_{\rm int}^{(\Lambda)}(\tilde{A})\Bigr]
\exp\bigl[-\frac{\beta}{M}\check{H}_{{\rm hop},\uparrow}^{(\Lambda)}(\tilde{A})\bigr]
\exp\bigl[-\frac{\beta}{M}\vartheta_{\rm spin}(\check{H}_{{\rm hop},\uparrow}^{(\Lambda)}(-\tilde{A}))\bigr].\ret
\end{eqnarray}
By using the expression (\ref{checkHint}) of the interaction Hamiltonian $\check{H}_{\rm int}^{(\Lambda)}(\tilde{A})$, 
we expand this right-hand side so that each term has the following form: 
\begin{eqnarray}
\label{IMspinexpand}
& &\left(\frac{\beta g}{M}\right)^\ell\exp[i\Theta]\times 
\check{K}_\uparrow \check{K}_\downarrow \cdots \check{K}_\uparrow \check{K}_\downarrow
\times a_{x_1,\uparrow}^\dagger a_{y_1,\uparrow}^\dagger \vartheta_{\rm spin}(a_{x_1,\uparrow}^\dagger a_{y_1,\uparrow}^\dagger)\ret
&\times&\check{K}_\uparrow \check{K}_\downarrow \cdots \check{K}_\uparrow \check{K}_\downarrow
\times a_{x_2,\uparrow}^\dagger a_{y_2,\uparrow}^\dagger \vartheta_{\rm spin}(a_{x_2,\uparrow}^\dagger a_{y_2,\uparrow}^\dagger)\ret
&\times& \cdots \times 
\check{K}_\uparrow \check{K}_\downarrow \cdots \check{K}_\uparrow \check{K}_\downarrow
\times a_{x_\ell,\uparrow}^\dagger a_{y_\ell,\uparrow}^\dagger 
\vartheta_{\rm spin}(a_{x_\ell,\uparrow}^\dagger a_{y_\ell,\uparrow}^\dagger)\times 
\check{K}_\uparrow \check{K}_\downarrow \cdots \check{K}_\uparrow \check{K}_\downarrow
\end{eqnarray}
with some integer $\ell\ge 0$, where we have written 
\begin{equation}
\check{K}_\uparrow:=\exp\bigl[-\frac{\beta}{M}\check{H}_{{\rm hop},\uparrow}^{(\Lambda)}(\tilde{A})\bigr]\quad 
\mbox{and} \quad \check{K}_\downarrow 
:=\exp\bigl[-\frac{\beta}{M}\vartheta_{\rm spin}(\check{H}_{{\rm hop},\uparrow}^{(\Lambda)}(-\tilde{A}))\bigr], 
\end{equation}
and $\Theta$ is the sum of the phases, each of which is the phase $\pm 2\tilde{A}_{x,y}$ in the interaction Hamiltonian 
$\check{H}_{\rm int}^{(\Lambda)}(\tilde{A})$. 
Since a fermion operator with even parity that consists of only spin up operators commutes with 
all the fermion operators that consist of only spin down operators, the term (\ref{IMspinexpand}) is written 
\begin{eqnarray}
\label{expandIMspineach}
& &\left(\frac{\beta g}{M}\right)^\ell\exp[i\Theta]\times \left(\check{K}_\uparrow\right)^{m_1}
a_{x_1,\uparrow}^\dagger a_{y_1,\uparrow}^\dagger\left(\check{K}_\uparrow\right)^{m_2}
a_{x_2,\uparrow}^\dagger a_{y_2,\uparrow}^\dagger \cdots \left(\check{K}_\uparrow\right)^{m_\ell}
a_{x_\ell,\uparrow}^\dagger a_{y_\ell,\uparrow}^\dagger\left(\check{K}_\uparrow\right)^{m_{\ell+1}}\ret
&\times&\left(\check{K}_\downarrow\right)^{m_1}
\vartheta_{\rm spin}(a_{x_1,\uparrow}^\dagger a_{y_1,\uparrow}^\dagger)\left(\check{K}_\downarrow\right)^{m_2}
\vartheta_{\rm spin}(a_{x_2,\uparrow}^\dagger a_{y_2,\uparrow}^\dagger)\cdots \left(\check{K}_\downarrow\right)^{m_\ell}
\vartheta_{\rm spin}(a_{x_\ell,\uparrow}^\dagger a_{y_\ell,\uparrow}^\dagger)\left(\check{K}_\downarrow\right)^{m_{\ell+1}},\ret
\end{eqnarray}
where $m_i$ are non-negative integers which satisfy 
$$
\sum_{i=1}^{\ell+1}m_i=M.
$$

In order to estimate the trace of the right-hand side of (\ref{IMspinM}), we recall some tools. 
We write $\mathfrak{A}_\sigma$ for the algebra generated by fermion operators, $a_{x,\sigma}$ and $a_{x',\sigma}^\dagger$,  
with a fixed spin $\sigma$ for $x,x'\in\Lambda$. Then, the following bound is valid: \cite{JP} 
\begin{equation}
{\rm Tr}\; \mathcal{A}\vartheta_{\rm spin}(\mathcal{A})\ge 0 \quad \mbox{for any } \mathcal{A}\in\mathfrak{A}_\uparrow.
\end{equation}
Let $\mathcal{A}_j\in\mathfrak{A}_\uparrow$ and $\mathcal{B}_j\in\mathfrak{A}_\uparrow$ be two sets, 
$\{\mathcal{A}_j\}$ and $\{\mathcal{B}_j\}$, of operators. We define an inner product for these two sets by 
\begin{equation}
\langle\!\langle \{\mathcal{A}_j\},\{\mathcal{B}_j\} \rangle\!\rangle_{\rm spin}:=\sum_j{\rm Tr}\; \mathcal{A}_j
\vartheta_{\rm spin}(\mathcal{B}_j).
\end{equation}
Then, by the Schwarz inequality, one has 
\begin{equation}
\left|\langle\!\langle \{\mathcal{A}_j\},\{\mathcal{B}_j\} \rangle\!\rangle_{\rm spin}\right|^2\le 
\langle\!\langle \{\mathcal{A}_j\},\{\mathcal{A}_j\}\rangle\!\rangle_{\rm spin}
\langle\!\langle \{\mathcal{B}_j\}, \{\mathcal{B}_j\} \rangle\!\rangle_{\rm spin}. 
\end{equation}
By combining this, (\ref{IMspinM}) and (\ref{expandIMspineach}), we obtain 
\begin{eqnarray}
\left\{{\rm Tr} \exp[-\beta\check{H}^{(\Lambda)}(\tilde{A})]\right\}^2&\le&
{\rm Tr}\exp[-\beta \check{H}_{{\rm hop},\uparrow}^{(\Lambda)}(\tilde{A})
-\beta\check{H}_{{\rm hop},\downarrow}^{(\Lambda)}(-\tilde{A})-\beta \check{H}_{\rm int}^{(\Lambda)}(0)]\ret
&\times& {\rm Tr}\exp[-\beta \check{H}_{{\rm hop},\uparrow}^{(\Lambda)}(-\tilde{A})
-\beta\check{H}_{{\rm hop},\downarrow}^{(\Lambda)}(\tilde{A})-\beta \check{H}_{\rm int}^{(\Lambda)}(0)].\ret 
\end{eqnarray}
Further, by using (\ref{checkHhop}) and (\ref{checkHint}), this can be written 
\begin{multline}
\label{Zbound}
\left\{{\rm Tr}\exp\bigl[-\beta\bigl(\bar{H}_{\rm hop}^{(\Lambda)}(\tilde{A})
+\bar{H}_{\rm int}^{(\Lambda)}(\tilde{A})\bigr)\bigr]\right\}^2\\
\le{\rm Tr} \exp\bigl[-\beta \bigl(\bar{H}_{{\rm hop},\uparrow}^{(\Lambda)}(\tilde{A})
+\bar{H}_{{\rm hop},\downarrow}^{(\Lambda)}(-\tilde{A})+\bar{H}_{\rm int}^{(\Lambda)}(0)\bigl)\bigr]\\
\times {\rm Tr}\exp\bigl[-\beta \bigl(\bar{H}_{{\rm hop},\uparrow}^{(\Lambda)}(-\tilde{A})
+\bar{H}_{{\rm hop},\downarrow}^{(\Lambda)}(\tilde{A})+\bar{H}_{\rm int}^{(\Lambda)}(0)\bigr)\bigr],
\end{multline}
where $\bar{H}_{{\rm hop},\sigma}^{(\Lambda)}(\tilde{A})$ is the spin-$\sigma$ part of $\bar{H}_{\rm hop}^{(\Lambda)}(\tilde{A})$. 

\Section{Reflection positivity --- real space}
\label{Sec:RPreal}

In order to further evaluate the right-hand side of (\ref{Zbound}), we rely on reflection positivity about real space \cite{JP}. 
Since the sidelengths of the cube $\Lambda$ of (\ref{Lambda}) are all the even integer $2L$, 
the lattice $\Lambda$ is invariant under a reflection $\vartheta$ in a plane $\Pi$ 
which is normal to a coordinate direction and intersects no sites in $\Lambda$, i.e., $\vartheta(\Lambda)=\Lambda$. 
Clearly, the lattice $\Lambda$ can be decomposed into two parts $\Lambda=\Lambda_-\cup\Lambda_+$, 
where $\Lambda_\pm$ denote the set of the sites on the $\pm$ side of the plane $\Pi$. 
The reflection $\vartheta$ maps $\Lambda_\pm$ into $\Lambda_\mp$, i.e., $\vartheta(\Lambda_\pm)=\Lambda_\mp$. 
Let $\Omega$ be a subset of $\Lambda$, i.e., $\Omega\subset\Lambda$. 
We write $\mathfrak{A}(\Omega)$ for the algebra generated by $a_{x,\sigma}$ and $a_{x',\sigma'}^\dagger$ for $x,x'\in\Omega$, 
$\sigma,\sigma'=\uparrow,\downarrow$. 
We also write $\mathfrak{A}=\mathfrak{A}(\Lambda)$, and $\mathfrak{A}_\pm=\mathfrak{A}(\Lambda_\pm)$. 

Following \cite{JP}, we consider an anti-linear representation of the reflection $\vartheta$ on the fermion 
Hilbert space, which is also denoted by $\vartheta$. 
The anti-linear map, $\vartheta:\mathfrak{A}_\pm\rightarrow\mathfrak{A}_\mp$, is defined by 
\begin{equation}
\label{defvartheta}
\vartheta(a_{x,\sigma})=a_{\vartheta(x),\sigma}\quad \mbox{and} \quad 
\vartheta(a_{x,\sigma}^\dagger)=a_{\vartheta(x),\sigma}^\dagger. 
\end{equation}
For $\mathcal{A},\mathcal{B}\in\mathfrak{A}$, 
\begin{equation}
\vartheta(\mathcal{A}\mathcal{B})=\vartheta(\mathcal{A})\vartheta(\mathcal{B})
\quad \mbox{and}\quad \vartheta(\mathcal{A})^\dagger=\vartheta(\mathcal{A}^\dagger). 
\end{equation}

We write  
\begin{equation}
\label{Hacute}
\acute{H}^{(\Lambda)}(\tilde{A}):=
\bar{H}_{{\rm hop},\uparrow}^{(\Lambda)}(\tilde{A})
+\bar{H}_{{\rm hop},\downarrow}^{(\Lambda)}(-\tilde{A})+\bar{H}_{\rm int}^{(\Lambda)}(0) 
\end{equation}
for the Hamiltonian in the right-hand side of (\ref{Zbound}). 
The interaction Hamiltonian $\bar{H}_{\rm int}^{(\Lambda)}(0)$ of (\ref{tildeHint}) with $\tilde{A}=0$ can be written \cite{Koma4}
\begin{equation}
\bar{H}_{\rm int}^{(\Lambda)}(0)=\sum_{j=1}^d \bar{H}_{{\rm int},j}^{(\Lambda)}
-\frac{dg}{2}\sum_{x\in\Lambda}\left\{[\Gamma_x^{(1)}]^2-[\Gamma_x^{(2)}]^2\right\},
\end{equation}
where 
\begin{equation}
\bar{H}_{{\rm int},j}^{(\Lambda)}:=\frac{g}{4}\sum_{x\in\Lambda}[\Gamma_x^{(1)}+\Gamma_{x+e_j}^{(1)}]^2
-\frac{g}{4}\sum_{x\in\Lambda}[\Gamma_x^{(2)}-\Gamma_{x+e_j}^{(2)}]^2,
\end{equation}
\begin{equation}
\label{Gamma}
\Gamma_x^{(1)}:=a_{x,\uparrow}^\dagger a_{x,\downarrow}^\dagger + a_{x,\downarrow}a_{x,\uparrow}
\quad \mbox{and} \quad 
\Gamma_x^{(2)}:=i(a_{x,\uparrow}^\dagger a_{x,\downarrow}^\dagger - a_{x,\downarrow}a_{x,\uparrow}).
\end{equation}
We also write 
\begin{equation}
\bar{H}_{{\rm hop},\sigma,i}^{(\Lambda)}(\tilde{A}):=i\kappa \sum_{x\in\Lambda}(-1)^{\theta_i(x)}
\bigl(a_{x,\sigma}^\dagger e^{i\eta_\sigma\tilde{A}_{x,x+e_i}}a_{x+e_i,\sigma}
-a_{x+e_i,\sigma}^\dagger e^{-i\eta_\sigma\tilde{A}_{x,x+e_i}}a_{x,\sigma}\bigr)
\end{equation}
for $i=1,2,\ldots,d$ and $\sigma=\uparrow,\downarrow$, where $\eta_\sigma$ is given by 
\begin{equation}
\eta_\sigma:=
\begin{cases}
1 & \mbox{for } \sigma=\uparrow\; ;\\
-1 & \mbox{for } \sigma=\downarrow. 
\end{cases}
\end{equation}
Then, 
\begin{equation}
\bar{H}_{{\rm hop},\uparrow}^{(\Lambda)}(\tilde{A})+\bar{H}_{{\rm hop},\downarrow}^{(\Lambda)}(-\tilde{A})
=\sum_{\sigma=\uparrow,\downarrow}\sum_{i=1}^d \bar{H}_{{\rm hop},\sigma,i}^{(\Lambda)}(\tilde{A}).
\end{equation}
Following \cite{Koma4}, we want to treat the Hamiltonian $\acute{H}^{(\Lambda)}(\tilde{A})$ of (\ref{Hacute}). 
Namely, we will apply the method of reflection positivity to the system.  

Since we have imposed the periodic boundary conditions for the lattice $\Lambda$, 
there are two planes which divide the lattice $\Lambda$ 
into the two halves, $\Lambda_-$ and $\Lambda_+$. We denote the two planes collectively by $\tilde{\Pi}$. 
Consider first the reflection in the plane $\Pi$ normal to the $x^{(1)}$ direction. 
Since we can change the locations of the bonds with the opposite sign of the hopping amplitudes 
due to the anti-periodic boundary conditions by using a gauge transformation as shown in \cite{Koma4}, 
we can take 
\begin{equation}
\Lambda_-=\{x|-L+1\le x^{(1)}\le 0\}\quad \mbox{and}\quad \Lambda_+=\{x|\; 1\le x^{(1)}\le L\}
\end{equation}
without loss of generality. 
Then, the bonds crossing the plane $\tilde{\Pi}$ are given by 
$$
\{(0,x^{(2)},\ldots,x^{(d)}),(1,x^{(2)},\ldots,x^{(d)})\}
$$ 
and 
$$
\{(-L+1,x^{(2)},\ldots,x^{(d)}),(L,x^{(2)},\ldots,x^{(d)})\}
$$ 
for $x^{(2)},\ldots,x^{(d)}\in\{-L+1,-L+2,\ldots,-1,0,1,\ldots,L-1,L\}$.

We introduce some unitary transformations as follows: 
\begin{equation}
\label{defU1j}
U_{1,j}^{(\Lambda)}
:=\prod_{\substack{x\in\Lambda,\; \sigma=\uparrow,\downarrow \\ :\;x^{(j)}={\rm even}}}e^{(i\pi/2)n_{x,\sigma}}
\quad \mbox{for \ } j=2,3,\ldots,d, 
\end{equation}
and 
\begin{equation}
\label{defU1}
U_1^{(\Lambda)}:=\prod_{j=2}^d U_{1,j}^{(\Lambda)}. 
\end{equation}
Since 
\begin{equation}
e^{-(i\pi/2)n_{x,\sigma}}a_{x,\sigma}e^{(i\pi/2)n_{x,\sigma}}=ia_{x,\sigma},
\end{equation}
we have 
\begin{equation}
(U_{1,j}^{(\Lambda)})^\dagger a_{x,\sigma}U_{1,j}^{(\Lambda)}=
\begin{cases}
ia_{x,\sigma}, &  \mbox{for}\ x^{(j)}={\rm even}; \\
a_{x,\sigma}, & \mbox{for}\ x^{(j)}={\rm odd}. 
\end{cases}
\end{equation}

Next, we introduce \cite{FILS} 
\begin{equation}
\label{uxsigma}
u_{x,\sigma}:=\left[\prod_{\substack{y\in\Lambda,\; \sigma'=\uparrow,\downarrow \\ :\; y\ne x\; {\rm or}\; \sigma'\ne \sigma}}
(-1)^{n_{y,\sigma'}}\right](a_{x,\sigma}^\dagger+a_{x,\sigma}). 
\end{equation}
Then, 
\begin{equation}
(u_{x,\sigma})^\dagger a_{y,\sigma'}u_{x,\sigma}=
\begin{cases}
a_{x,\sigma}^\dagger, & \mbox{for}\; y=x \ \mbox{and} \ \sigma'=\sigma;\\
a_{y,\sigma'}, & \mbox{otherwise}.
\end{cases}
\end{equation}
By using these operators, we further introduce \cite{FILS} 
\begin{equation}
\label{defU2}
U_{\rm odd}^{(\Lambda)}:=\prod_{\sigma=\uparrow,\downarrow}\prod_{x\in\Lambda_{\rm odd}}u_{x,\sigma}.
\end{equation}
Immediately, 
\begin{equation}
\label{U2}
(U_{\rm odd}^{(\Lambda)})^\dagger a_{x,\sigma}U_{\rm odd}^{(\Lambda)}=
\begin{cases}
a_{x,\sigma}^\dagger, & \mbox{for}\; x\in \Lambda_{\rm odd};\\
a_{x,\sigma}, & \mbox{otherwise}.  
\end{cases}
\end{equation}
We write 
\begin{equation}
\label{U}
\tilde{U}_1^{(\Lambda)}:=U_1^{(\Lambda)}U_{\rm odd}^{(\Lambda)}.
\end{equation}

In the same way as in \cite{Koma4}, one has 
\begin{equation}
\tilde{\bar{H}}_{{\rm int},1}^{(\Lambda)}
:=(\tilde{U}_1^{(\Lambda)})^\dagger \bar{H}_{{\rm int},1}^{(\Lambda)}\tilde{U}_1^{(\Lambda)}
=\frac{g}{4}\sum_{x\in\Lambda}[\Gamma_x^{(1)}-\Gamma_{x+e_1}^{(1)}]^2 
-\frac{g}{4}\sum_{x\in\Lambda}[\Gamma_x^{(2)}-\Gamma_{x+e_1}^{(2)}]^2, 
\end{equation}
and 
\begin{equation}
\tilde{\bar{H}}_{{\rm int},j}^{(\Lambda)}:=
(\tilde{U}_1^{(\Lambda)})^\dagger \bar{H}_{{\rm int},j}^{(\Lambda)}\tilde{U}_1^{(\Lambda)}
=\frac{g}{4}\sum_{x\in\Lambda}[\Gamma_x^{(1)}+\Gamma_{x+e_j}^{(1)}]^2 
-\frac{g}{4}\sum_{x\in\Lambda}[\Gamma_x^{(2)}+\Gamma_{x+e_j}^{(2)}]^2
\end{equation}
for $j=2,3,\ldots,d$. We write 
\begin{equation}
\tilde{\bar{H}}_{\rm int}^{(\Lambda)}(0)
:=(\tilde{U}_1^{(\Lambda)})^\dagger \bar{H}_{\rm int}^{(\Lambda)}(0)\tilde{U}_1^{(\Lambda)}.
\end{equation}
Then, from the above observations, this Hamiltonian can be decomposed into three parts, 
\begin{equation}
\label{decomptildebarHint}
\tilde{\bar{H}}_{\rm int}^{(\Lambda)}(0)=\tilde{\bar{H}}_{{\rm int},-}^{(\Lambda)}+\tilde{\bar{H}}_{{\rm int},+}^{(\Lambda)}
+\tilde{\bar{H}}_{{\rm int},0}^{(\Lambda)},
\end{equation}
where the two Hamiltonians $\tilde{\bar{H}}_{{\rm int},\pm}^{(\Lambda)}$ satisfy 
$\tilde{\bar{H}}_{{\rm int},\pm}^{(\Lambda)}\in\mathfrak{A}_\pm$ and 
$\vartheta(\tilde{\bar{H}}_{{\rm int},-}^{(\Lambda)})=\tilde{\bar{H}}_{{\rm int},+}^{(\Lambda)}$, and 
the third Hamiltonian is given by 
\begin{eqnarray}
\label{tildebarHint0}
\tilde{\bar{H}}_{{\rm int},0}^{(\Lambda)}&:=&\frac{g}{4}\sum_{x^{(2)},\ldots,x^{(d)}}[\Gamma_{x_1^0}^{(1)}-\Gamma_{x_1^1}^{(1)}]^2
+\frac{g}{4}\sum_{x^{(2)},\ldots,x^{(d)}}[\Gamma_{x_1^+}^{(1)}-\Gamma_{x_1^-}^{(1)}]^2\ret
&-&\frac{g}{4}\sum_{x^{(2)},\ldots,x^{(d)}}[\Gamma_{x_1^0}^{(2)}-\Gamma_{x_1^1}^{(2)}]^2
-\frac{g}{4}\sum_{x^{(2)},\ldots,x^{(d)}}[\Gamma_{x_1^+}^{(2)}-\Gamma_{x_1^-}^{(2)}]^2,
\end{eqnarray}
where we have written 
\begin{equation}
x_1^0:=(0,x^{(2)},\ldots,x^{(d)}),\quad x_1^1:=(1,x^{(2)},\ldots,x^{(d)}),
\end{equation}
and 
\begin{equation} 
x_1^+:=(L,x^{(2)},\ldots,x^{(d)}),\quad x_1^-:=(-L+1,x^{(2)},\ldots,x^{(d)}).
\end{equation}
Similarly, one has 
\begin{eqnarray}
\label{tildebarHhopsigma1}
\tilde{\bar{H}}_{{\rm hop},\sigma,1}^{(\Lambda)}(\tilde{A})&:=&
(\tilde{U}_1^{(\Lambda)})^\dagger\bar{H}_{{\rm hop},\sigma,1}^{(\Lambda)}(\tilde{A})\tilde{U}_1^{(\Lambda)}\ret
&=&i\kappa \sum_{x\in\Lambda}(-1)^{\theta_1(x)}
\bigl[e^{i\upsilon_\sigma(x)\tilde{A}_{x,x+e_1}}a_{x,\sigma}^\dagger a_{x+e_1,\sigma}^\dagger
+e^{-i\upsilon_\sigma(x)\tilde{A}_{x,x+e_1}}a_{x,\sigma}a_{x+e_1,\sigma}\bigr],\ret
\end{eqnarray}
where we have written $\upsilon_\sigma(x):=(-1)^{x^{(1)}+\cdots+x^{(d)}}\eta_\sigma$. 
This Hamiltonian can be decomposed into three parts, 
\begin{equation}
\label{tildebarHhopdecomp}
\tilde{\bar{H}}_{{\rm hop},\sigma,1}^{(\Lambda)}(\tilde{A})=\tilde{\bar{H}}_{{\rm hop},\sigma,1,-}^{(\Lambda)}(\tilde{A})
+\tilde{\bar{H}}_{{\rm hop},\sigma,1,+}^{(\Lambda)}(\tilde{A})
+\tilde{\bar{H}}_{{\rm hop},\sigma,1,0}^{(\Lambda)}(\tilde{A})
\end{equation}
where $\tilde{\bar{H}}_{{\rm hop},\sigma,1,\pm}^{(\Lambda)}(\tilde{A})\in\mathfrak{A}_\pm$ and the third part is given by 
\begin{multline}
\tilde{\bar{H}}_{{\rm hop},\sigma,1,0}^{(\Lambda)}(\tilde{A}):=
i\kappa \sum_{x^{(2)},\ldots,x^{(d)}}
\bigl\{\exp[{i\upsilon_\sigma(x)\tilde{A}_{x_1^0,x_1^1}}]
a_{x_1^0,\sigma}^\dagger a_{x_1^1,\sigma}^\dagger+ \exp[{-i\upsilon_\sigma(x)\tilde{A}_{x_1^0,x_1^1}}]
a_{x_1^0,\sigma}a_{x_1^1,\sigma}\bigr\}\\
-i\kappa \sum_{x^{(2)},\ldots,x^{(d)}}
\bigl\{\exp[{i\upsilon_\sigma(x)\tilde{A}_{x_1^+,x_1^-}}]
a_{x_1^+,\sigma}^\dagger a_{x_1^-,\sigma}^\dagger+ \exp[{-i\upsilon_\sigma(x)\tilde{A}_{x_1^+,x_1^-}}]
a_{x_1^+,\sigma}a_{x_1^-,\sigma}\bigr\}.
\end{multline}
By using the reflection $\vartheta$, this can be written as 
\begin{eqnarray}
\label{tildebarHhopsigma10}
& &\hspace{-1.3cm}\tilde{\bar{H}}_{{\rm hop},\sigma,1,0}^{(\Lambda)}(\tilde{A})\ret
&=&i\kappa \sum_{x^{(2)},\ldots,x^{(d)}}
\bigl\{\exp[{i\upsilon_\sigma(x)\tilde{A}_{x_1^0,x_1^1}}]
a_{x_1^0,\sigma}^\dagger \vartheta(a_{x_1^0,\sigma}^\dagger)+ \exp[{-i\upsilon_\sigma(x)\tilde{A}_{x_1^0,x_1^1}}]
a_{x_1^0,\sigma}\vartheta(a_{x_1^0,\sigma})\bigr\}\ret
&+&i\kappa \sum_{x^{(2)},\ldots,x^{(d)}}
\bigl\{\exp[{i\upsilon_\sigma(x)\tilde{A}_{x_1^+,x_1^-}}]
a_{x_1^-,\sigma}^\dagger \vartheta(a_{x_1^-,\sigma}^\dagger)+ \exp[{-i\upsilon_\sigma(x)\tilde{A}_{x_1^+,x_1^-}}]
a_{x_1^-,\sigma}\vartheta(a_{x_1^-,\sigma})\bigr\}.\ret
\end{eqnarray}
Further, one has 
\begin{eqnarray}
\label{tildebarHhopsigmai}
\tilde{\bar{H}}_{{\rm hop},\sigma,i}^{(\Lambda)}(\tilde{A})&:=&
(\tilde{U}_1^{(\Lambda)})^\dagger\bar{H}_{{\rm hop},\sigma,i}^{(\Lambda)}(\tilde{A})\tilde{U}_1^{(\Lambda)}\ret
&=&\kappa \sum_{x\in\Lambda}\varrho_i(x)\upsilon(x)
\bigl[e^{i\upsilon_\sigma(x)\tilde{A}_{x,x+e_i}}a_{x,\sigma}^\dagger a_{x+e_i,\sigma}^\dagger
+e^{-i\upsilon_\sigma(x)\tilde{A}_{x,x+e_i}}a_{x+e_i,\sigma}a_{x,\sigma}\bigr]\ret
\end{eqnarray}
for $i=2,3,\ldots,d$, where we have written $\varrho_i(x):=(-1)^{\theta_i(x)+x^{(i)}}$ and $\upsilon(x):=(-1)^{x^{(1)}+\cdots+x^{(d)}}$. 
Clearly, each of these Hamiltonians can be decomposed into two parts, 
\begin{equation}
\tilde{\bar{H}}_{{\rm hop},\sigma,i}^{(\Lambda)}(\tilde{A})=\tilde{\bar{H}}_{{\rm hop},\sigma,i,-}^{(\Lambda)}(\tilde{A})
+\tilde{\bar{H}}_{{\rm hop},\sigma,i,+}^{(\Lambda)}(\tilde{A}), 
\end{equation}
where $\tilde{\bar{H}}_{{\rm hop},\sigma,i,\pm}^{(\Lambda)}(\tilde{A})$ satisfy 
$\tilde{\bar{H}}_{{\rm hop},\sigma,i,\pm}^{(\Lambda)}(\tilde{A})\in\mathfrak{A}_\pm$.

Now let us consider the Hamiltonian, 
\begin{eqnarray}
\label{defgraveH}
\grave{H}^{(\Lambda)}(\tilde{A})&:=&(\tilde{U}_1^{(\Lambda)})^\dagger\acute{H}^{(\Lambda)}(\tilde{A})\tilde{U}_1^{(\Lambda)}\ret
&=&(\tilde{U}_1^{(\Lambda)})^\dagger\bigl[
\bar{H}_{{\rm hop},\uparrow}^{(\Lambda)}(\tilde{A})
+\bar{H}_{{\rm hop},\downarrow}^{(\Lambda)}(-\tilde{A})+\bar{H}_{\rm int}^{(\Lambda)}(0)\bigr]\tilde{U}_1^{(\Lambda)}\ret
&=& \tilde{\bar{H}}_{{\rm hop},\uparrow}^{(\Lambda)}(\tilde{A})+\tilde{\bar{H}}_{{\rm hop},\downarrow}^{(\Lambda)}(\tilde{A})
+\tilde{\bar{H}}_{\rm int}^{(\Lambda)}(0),
\end{eqnarray} 
where we have written
\begin{equation}
\tilde{\bar{H}}_{{\rm hop},\sigma}^{(\Lambda)}(\tilde{A})
:=\sum_{i=1}^d \tilde{\bar{H}}_{{\rm hop},\sigma,i}^{(\Lambda)}(\tilde{A})\quad \mbox{for \ } \sigma=\uparrow,\downarrow.
\end{equation}
In the same way, this Hamiltonian can be written 
\begin{equation}
\grave{H}^{(\Lambda)}(\tilde{A})=\grave{H}_-^{(\Lambda)}(\tilde{A})+\grave{H}_+^{(\Lambda)}(\tilde{A})
+\grave{H}_0^{(\Lambda)}(\tilde{A}),
\end{equation}
where $\grave{H}_\pm^{(\Lambda)}(\tilde{A})$ satisfy $\grave{H}_\pm^{(\Lambda)}(\tilde{A})\in\mathfrak{A}_\pm$. 
{From} (\ref{decomptildebarHint}), (\ref{tildebarHint0}), (\ref{tildebarHhopdecomp}) and (\ref{tildebarHhopsigma10}), 
the explicit form of the third term in the right-hand side is given by 
\begin{equation}
\grave{H}_0^{(\Lambda)}(\tilde{A})=\grave{H}_{0,{\rm hop}}^{(\Lambda)}(\tilde{A})
+\grave{H}_{0,{\rm int}}^{(\Lambda)}
\end{equation}
with
\begin{eqnarray}
& &\hspace{-1.1cm}\grave{H}_{0,{\rm hop}}^{(\Lambda)}(\tilde{A})\ret
&:=&i\kappa \sum_{\sigma=\uparrow,\downarrow}\sum_{x^{(2)},\ldots,x^{(d)}}
\bigl\{\exp[{i\upsilon_\sigma(x)\tilde{A}_{x_1^0,x_1^1}}]
a_{x_1^0,\sigma}^\dagger \vartheta(a_{x_1^0,\sigma}^\dagger)+ \exp[{i\upsilon_\sigma(x)\tilde{A}_{x_1^1,x_1^0}}]
a_{x_1^0,\sigma}\vartheta(a_{x_1^0,\sigma})\bigr\}\ret
&+&i\kappa \sum_{\sigma=\uparrow,\downarrow}\sum_{x^{(2)},\ldots,x^{(d)}}
\bigl\{\exp[{i\upsilon_\sigma(x)\tilde{A}_{x_1^+,x_1^-}}]
a_{x_1^-,\sigma}^\dagger \vartheta(a_{x_1^-,\sigma}^\dagger)+ \exp[{i\upsilon_\sigma(x)\tilde{A}_{x_1^-,x_1^+}}]
a_{x_1^-,\sigma}\vartheta(a_{x_1^-,\sigma})\bigr\}\ret
\end{eqnarray}
and
\begin{eqnarray}
\grave{H}_{0,{\rm int}}^{(\Lambda)}
&:=&\frac{g}{4}\sum_{x^{(2)},\ldots,x^{(d)}}[\Gamma_{x_1^0}^{(1)}-\Gamma_{x_1^1}^{(1)}]^2
+\frac{g}{4}\sum_{x^{(2)},\ldots,x^{(d)}}[\Gamma_{x_1^+}^{(1)}-\Gamma_{x_1^-}^{(1)}]^2\ret
&-&\frac{g}{4}\sum_{x^{(2)},\ldots,x^{(d)}}[\Gamma_{x_1^0}^{(2)}-\Gamma_{x_1^1}^{(2)}]^2
-\frac{g}{4}\sum_{x^{(2)},\ldots,x^{(d)}}[\Gamma_{x_1^+}^{(2)}-\Gamma_{x_1^-}^{(2)}]^2.
\end{eqnarray}
Therefore, in the same way as in \cite{Koma4}, we obtain 
\begin{eqnarray}
\label{TrexpgraveHbound}
\left\{{\rm Tr} \exp[-\beta \grave{H}^{(\Lambda)}(\tilde{A})]\right\}^2
&\le& {\rm Tr} \exp[-\beta \grave{H}_-^{(\Lambda)}(\tilde{A})-\beta\vartheta(\grave{H}_-^{(\Lambda)}(\tilde{A}))
-\beta \grave{H}_0^{(\Lambda)}(0)]\ret
&\times& {\rm Tr} \exp[-\beta \vartheta(\grave{H}_+^{(\Lambda)}(\tilde{A}))-\beta\grave{H}_+^{(\Lambda)}(\tilde{A}) 
-\beta \grave{H}_0^{(\Lambda)}(0)]. 
\end{eqnarray}
{From} (\ref{tildebarHhopsigma1}) and (\ref{tildebarHhopsigmai}), 
the explicit form of the Hamiltonian $\grave{H}_-^{(\Lambda)}(\tilde{A})$ is given by 
\begin{eqnarray}
& &\hspace{-1.2cm}\grave{H}_-^{(\Lambda)}(\tilde{A})\ret
&=&i\kappa \sum_{\sigma=\uparrow,\downarrow}
\sum_{\substack{x\in\Lambda_- \\ :\; x+e_1\in\Lambda_-}}
\bigl\{\exp[i\upsilon_\sigma(x)\tilde{A}_{x,x+e_1}]a_{x,\sigma}^\dagger a_{x+e_1,\sigma}^\dagger 
+\exp[-i\upsilon_\sigma(x)\tilde{A}_{x,x+e_1}]a_{x,\sigma}a_{x+e_1,\sigma}\bigr\}\ret
&+&\kappa \sum_{\sigma=\uparrow,\downarrow}\sum_{i=2}^d \sum_{x\in\Lambda_-}\tilde{\varrho}_i(x)
\bigl\{\exp[i\upsilon_\sigma(x)\tilde{A}_{x,x+e_i}]a_{x,\sigma}^\dagger a_{x+e_i,\sigma}^\dagger 
+\exp[-i\upsilon_\sigma(x)\tilde{A}_{x,x+e_i}]a_{x+e_i,\sigma}a_{x,\sigma}\bigr\}\ret
&+&\tilde{\bar{H}}_{{\rm int},-}^{(\Lambda)}, 
\end{eqnarray}
where we have written $\tilde{\varrho}_i(x):=\varrho_i(x)\upsilon(x)$. Therefore, we have 
\begin{multline}
\vartheta(\grave{H}_-^{(\Lambda)}(\tilde{A}))
=-i\kappa \sum_{\sigma=\uparrow,\downarrow}
\sum_{\substack{x\in\Lambda_- \\ :\; x+e_1\in\Lambda_-}}
\bigl\{\exp[i\upsilon_\sigma(\vartheta(x))\tilde{A}_{x,x+e_1}]a_{\vartheta(x),\sigma}^\dagger a_{\vartheta(x+e_1),\sigma}^\dagger\\ 
+\exp[-i\upsilon_\sigma(\vartheta(x))\tilde{A}_{x,x+e_1}]a_{\vartheta(x),\sigma}a_{\vartheta(x+e_1),\sigma}\bigr\}\\
+\kappa \sum_{\sigma=\uparrow,\downarrow}\sum_{i=2}^d \sum_{x\in\Lambda_-}\tilde{\varrho}_i(\vartheta(x))
\bigl\{\exp[i\upsilon_\sigma(\vartheta(x))\tilde{A}_{x,x+e_i}]a_{\vartheta(x),\sigma}^\dagger a_{\vartheta(x+e_i),\sigma}^\dagger\\ 
+\exp[-i\upsilon_\sigma(\vartheta(x))\tilde{A}_{x,x+e_i}]a_{\vartheta(x+e_i),\sigma}a_{\vartheta(x),\sigma}\bigr\}
+\tilde{\bar{H}}_{{\rm int},+}^{(\Lambda)}\\
=i\kappa \sum_{\sigma=\uparrow,\downarrow}
\sum_{\substack{x\in\Lambda_- \\ :\; x+e_1\in\Lambda_-}}
\bigl\{\exp[i\upsilon_\sigma(\vartheta(x))\tilde{A}_{x,x+e_1}]a_{\vartheta(x+e_1),\sigma}^\dagger a_{\vartheta(x),\sigma}^\dagger \\ 
+\exp[-i\upsilon_\sigma(\vartheta(x))\tilde{A}_{x,x+e_1}]a_{\vartheta(x+e_1),\sigma}a_{\vartheta(x),\sigma}\bigr\}\\
+\kappa \sum_{\sigma=\uparrow,\downarrow}\sum_{i=2}^d \sum_{x\in\Lambda_-}\tilde{\varrho}_i(\vartheta(x))
\bigl\{\exp[i\upsilon_\sigma(\vartheta(x))\tilde{A}_{x,x+e_i}]a_{\vartheta(x),\sigma}^\dagger a_{\vartheta(x+e_i),\sigma}^\dagger\\ 
+\exp[-i\upsilon_\sigma(\vartheta(x))\tilde{A}_{x,x+e_i}]a_{\vartheta(x+e_i),\sigma}a_{\vartheta(x),\sigma}\bigr\}
+\tilde{\bar{H}}_{{\rm int},+}^{(\Lambda)} 
\end{multline}
where we have used $\upsilon_\sigma(\vartheta(x))=-\upsilon_\sigma(x)$, $\tilde{\varrho}(\vartheta(x))=\tilde{\varrho}(x)$ and 
$\vartheta(\tilde{\bar{H}}_{{\rm int},-}^{(\Lambda)})=\tilde{\bar{H}}_{{\rm int},+}^{(\Lambda)}$. 
This can be rewritten as 
\begin{equation}
\vartheta(\grave{H}_-^{(\Lambda)}(\tilde{A}))=\vartheta(\grave{H}_-^{(\Lambda)}(\tilde{A}^-))=\grave{H}_+^{(\Lambda)}(\tilde{A}^-),
\end{equation}
where the gauge field $\tilde{A}^-$ is given by 
\begin{equation}
\tilde{A}_{x,y}^-:=\tilde{A}_{x,y} \quad \mbox{for } x,y\in\Lambda_-,
\end{equation}
\begin{equation}
\tilde{A}_{y,y+e_1}^-:=\tilde{A}_{\vartheta(y+e_1),\vartheta(y)}\quad \mbox{for } \{y,y+e_1\}\subset\Lambda_+,
\end{equation}
and 
\begin{equation}
\tilde{A}_{y,y+e_i}^-:=\tilde{A}_{\vartheta(y),\vartheta(y+e_i)}\quad \mbox{for } y\in\Lambda_+, \ 
i=2,3,\ldots,d.
\end{equation}
In the same way, we have 
\begin{equation}
\vartheta(\grave{H}_+^{(\Lambda)}(\tilde{A})=\vartheta(\grave{H}_+^{(\Lambda)}(\tilde{A}^+)
=\grave{H}_-^{(\Lambda)}(\tilde{A}^+),
\end{equation}
where the gauge field $\tilde{A}^+$ is defined by using $\tilde{A}$ on $\Lambda_+$. Substituting these results into 
the right-hand side of (\ref{TrexpgraveHbound}), we obtain 
\begin{eqnarray}
\left\{{\rm Tr} \exp[-\beta \grave{H}^{(\Lambda)}(\tilde{A})]\right\}^2
&\le& {\rm Tr} \exp[-\beta \grave{H}_-^{(\Lambda)}(\tilde{A}^-)-\beta\grave{H}_+^{(\Lambda)}(\tilde{A}^-)
-\beta \grave{H}_0^{(\Lambda)}(0)]\ret
&\times& {\rm Tr} \exp[-\beta \grave{H}_-^{(\Lambda)}(\tilde{A}^+)-\beta\grave{H}_+^{(\Lambda)}(\tilde{A}^+) 
-\beta \grave{H}_0^{(\Lambda)}(0)]. 
\end{eqnarray}
Further, since we can take $\tilde{A}_{x,y}^\pm$ to be equal to zero on the bonds $\{x,y\}$ 
crossing the reflection plane $\tilde{\Pi}$, this inequality can be rewritten as 
\begin{equation}
\left\{{\rm Tr} \exp[-\beta \grave{H}^{(\Lambda)}(\tilde{A})]\right\}^2
\le {\rm Tr} \exp[-\beta \grave{H}^{(\Lambda)}(\tilde{A}^-)]\times {\rm Tr} \exp[-\beta \grave{H}^{(\Lambda)}(\tilde{A}^+)].
\end{equation}
By using the relation (\ref{defgraveH}) between $\grave{H}^{(\Lambda)}(\tilde{A})$ and $\acute{H}^{(\Lambda)}(\tilde{A})$, 
this yields 
\begin{equation}
\left\{{\rm Tr} \exp[-\beta \acute{H}^{(\Lambda)}(\tilde{A})]\right\}^2
\le {\rm Tr} \exp[-\beta \acute{H}^{(\Lambda)}(\tilde{A}^-)]\times {\rm Tr} \exp[-\beta \acute{H}^{(\Lambda)}(\tilde{A}^+)].
\end{equation}
As shown in the previous paper \cite{Koma4}, we can interchange the roles of the
hopping amplitudes in the $x^{(1)}$ and $x^{(j)}$ directions for all $j=2,3,\ldots,d$. Combining these
observations with the argument in the proof of Theorem~4.2 in \cite{DLS}, we obtain the desired bound, 
\begin{equation}
{\rm Tr} \exp[-\beta \acute{H}^{(\Lambda)}(\tilde{A})]
\le {\rm Tr} \exp[-\beta \acute{H}^{(\Lambda)}(0)]. 
\end{equation}
Namely, the gauge field $\tilde{A}=0$ gives the minimum energy. Further, from (\ref{Zbound}) and (\ref{Hacute}), we obtain: 

\begin{theorem}
\label{maintheorem} 
The following inequality is valid for any gauge field $\tilde{A}$: 
\begin{equation}
{\rm Tr}\exp\bigl[-\beta\bigl(\bar{H}_{\rm hop}^{(\Lambda)}(\tilde{A})
+\bar{H}_{\rm int}^{(\Lambda)}(\tilde{A})\bigr)\bigr]\le 
{\rm Tr}\exp\bigl[-\beta\bigl(\bar{H}_{\rm hop}^{(\Lambda)}(0)
+\bar{H}_{\rm int}^{(\Lambda)}(0)\bigr)\bigr]. 
\end{equation}
\end{theorem}
\medskip

\noindent
Clearly, this inequality implies that the gauge field $\tilde{A}=0$ also minimizes the ground-state energy. 
In other words, the $\pi$ flux minimizes the ground-state energy \cite{LiebFlux}.

\Section{Absence of long-range order and existence of string long-range order}
\label{Sec:LRO}

As proved in the preceding section, the $\pi$ flux minimizes the ground-state energy of the full Hamiltonian of (\ref{fullbarHtildeA}). 
However, there are many configurations of the gauge fields $\tilde{A}$ that are gauge equivalent to $\tilde{A}=0$. 
As is well known, there are at least two ways to handle the gauge degrees of freedom: One is to fix the degrees of freedom, 
the other is to take account of all the configurations. 

In order to examine the two cases, let us consider the ground-state expectation value, 
\begin{equation}
\omega_0^{(\Lambda)}(\cdots)(\tilde{A}):=\lim_{\beta\nearrow\infty}\frac{1}{Z_\beta^{(\Lambda)}(\tilde{A})}{\rm Tr}(\cdots)
\exp\bigl[-\beta\bigl(\bar{H}_{\rm hop}^{(\Lambda)}(\tilde{A})+\bar{H}_{\rm int}^{(\Lambda)}(\tilde{A})\bigr)\bigr],
\end{equation}
for a given gauge field $\tilde{A}$, where 
$$
Z_\beta^{(\Lambda)}(\tilde{A}):={\rm Tr}
\exp\bigl[-\beta\bigl(\bar{H}_{\rm hop}^{(\Lambda)}(\tilde{A})+\bar{H}_{\rm int}^{(\Lambda)}(\tilde{A})\bigr)\bigr]. 
$$

\subsection{Case of gauge fixing}

As proved in our previous paper \cite{Koma4}, the ground state shows the long-range order for superconductivity 
when $\tilde{A}=0$ and $d\ge 3$. More precisely, the superconducting correlation $\omega_0(\Gamma_x^{(1)}\Gamma_y^{(1)})(0)$ 
does not decay with large distance, where $\Gamma_x^{(1)}$ are given by (\ref{Gamma}), and 
we have written 
\begin{equation}
\omega_0(\cdots)(0):={\rm weak}{-}\lim_{\Lambda\nearrow\ze^d}\omega_0^{(\Lambda)}(\cdots)(0) 
\end{equation}
for the infinite-volume ground state. In this case, the U(1) symmetry breaking occurs, and the corresponding Nambu-Goldstone 
mode appears above the symmetry-breaking ground state \cite{Koma4}. 

\subsection{Case without gauge fixing}

{From} the expression (\ref{Gamma}) of $\Gamma_x^{(1)}$, one has 
\begin{eqnarray}
\omega_0^{(\Lambda)}(\Gamma_x^{(1)}\Gamma_y^{(1)})(\tilde{A})
&=&\omega_0^{(\Lambda)}(a_{x,\uparrow}^\dagger a_{x,\downarrow}^\dagger a_{y,\downarrow}a_{y,\uparrow})(\tilde{A})
+\omega_0^{(\Lambda)}(a_{x,\downarrow}a_{x,\uparrow}a_{y,\uparrow}^\dagger a_{y,\downarrow}^\dagger)(\tilde{A})\ret
&=&2\times {\rm Re}\;\omega_0^{(\Lambda)}(a_{x,\uparrow}^\dagger a_{x,\downarrow}^\dagger a_{y,\downarrow}a_{y,\uparrow})(\tilde{A})
\end{eqnarray}
for $x\ne y$, where Re denotes the real part of the quantity, and we have used the conservation of the total number of fermions. 

Consider the class of the gauge fields which are gauge equivalent to $\tilde{A}=0$. 
These gauge fields are written as  
\begin{equation}
\label{tildeAvarphi}
\tilde{A}_{x,y}=\varphi_x-\varphi_y,
\end{equation}
where $\varphi_x\in(-\pi,\pi]$ is a function of the site $x\in\Lambda$. We introduce a ground-state expectation 
averaged over the class that is defined by 
\begin{equation}
\omega_0^{(\Lambda)}(\cdots):=\int d\varphi\; \omega_0^{(\Lambda)}(\cdots)(\tilde{A}),
\end{equation}
where we have written 
$$
\int d\varphi:=\prod_{x\in\Lambda}\int_{-\pi}^{\pi}\frac{d\varphi_x}{2\pi}.
$$

Let us consider the superconducting correlation 
$\omega_0^{(\Lambda)}(a_{x,\uparrow}^\dagger a_{x,\downarrow}^\dagger a_{y,\downarrow}a_{y,\uparrow})$. 
In order to calculate the value of this correlation, we introduce a unitary transformation by 
\begin{equation}
U^{(\Lambda)}(\varphi):=\prod_{\sigma=\uparrow,\downarrow}\prod_{x\in\Lambda}e^{i\varphi_x n_{x,\sigma}}. 
\end{equation}
Then, one has 
\begin{equation}
[U^{(\Lambda)}(\varphi)]^\dagger a_{x,\sigma}U^{(\Lambda)}(\varphi)=e^{i\varphi_x}a_{x,\sigma}
\end{equation}
for $x\in\Lambda$ and $\sigma=\uparrow,\downarrow$. By using (\ref{tildeAvarphi}), one has 
\begin{equation}
[U^{(\Lambda)}(\varphi)]^\dagger a_{x,\sigma}^\dagger e^{i\tilde{A}_{x,y}}a_{y,\sigma}U^{(\Lambda)}(\varphi)
=a_{x,\sigma}^\dagger a_{y,\sigma}.
\end{equation}
This yields 
\begin{equation}
[U^{(\Lambda)}(\varphi)]^\dagger [\bar{H}_{\rm hop}^{(\Lambda)}(\tilde{A})+\bar{H}_{\rm int}^{(\Lambda)}(\tilde{A})]
U^{(\Lambda)}(\varphi)=\bar{H}_{\rm hop}^{(\Lambda)}(0)+\bar{H}_{\rm int}^{(\Lambda)}(0). 
\end{equation}
By using this equality, we have 
\begin{eqnarray}
\omega_0^{(\Lambda)}(a_{x,\uparrow}^\dagger a_{x,\downarrow}^\dagger a_{y,\downarrow}a_{y,\uparrow})
&:=&\int d\varphi\; \omega_0^{(\Lambda)}(a_{x,\uparrow}^\dagger a_{x,\downarrow}^\dagger a_{y,\downarrow}a_{y,\uparrow})(\tilde{A})\ret
&=&\int d\varphi\; \omega_0^{(\Lambda)}(a_{x,\uparrow}^\dagger a_{x,\downarrow}^\dagger a_{y,\downarrow}a_{y,\uparrow})(0)
\times e^{-2i\varphi_x}e^{2i\varphi_y}\ret
&=&0 
\end{eqnarray}
for $x\ne y$. Thus, the superconducting correlation is vanishing when averaged over the gauge-equivalent class to $\tilde{A}=0$. 

Next, let us consider the gauge-invariant correlations. Let $\gamma$ be a path from $x$ to $y$ with the length $\ell$. 
Namely, there exist $\ell+1$ sites, $x=x_0, x_1,x_2,\ldots,x_{\ell-1},x_\ell=y$, 
such that $|x_j-x_{j+1}|=1$ for $j=0,1,\ldots,\ell-1$. We write 
\begin{equation}
\tilde{A}[\gamma]:=\sum_{j=0}^{\ell-1} \tilde{A}_{x_j,x_{j+1}}. 
\end{equation}
The string observable for superconductivity is defined by 
\begin{equation}
a_{x,\uparrow}^\dagger a_{x,\downarrow}^\dagger e^{2i\tilde{A}[\gamma]}a_{y,\downarrow}a_{y,\uparrow},
\end{equation}
which is gauge invariant. Then, in the same way as above, we obtain 
\begin{equation}
\omega_0^{(\Lambda)}(a_{x,\uparrow}^\dagger e^{2i\tilde{A}[\gamma]}a_{x,\downarrow}^\dagger a_{y,\downarrow}a_{y,\uparrow})
=\omega_0^{(\Lambda)}(a_{x,\uparrow}^\dagger a_{x,\downarrow}^\dagger a_{y,\downarrow}a_{y,\uparrow})(0). 
\end{equation}
Since the right-hand side is nothing but the superconducting correlation with the gauge field fixed to $\tilde{A}=0$, 
the string correlation shows the same long-range order.


\end{document}